# On the magnetism of σ-Fe$_{54}$Cr$_{46}$ alloy: AC and DC susceptibility studies

S. M. Dubiel[*], AGH University of Science and Technology, 30-059 Krakow, Poland

M.I. Tsindlekht and I. Felner, Racah Institute of Physics, The Hebrew University, Jerusalem, Israel 91904

Abstract

Sigma-phase intermetallic compound of Fe$_{54.2}$Cr$_{45.8}$ was investigated using DC and AC magnetic susceptibility techniques. A clear-cut evidence was found that the sample orders magnetically at $T_C$=23.5 K and its ground magnetic state is constituted by a spin glass. The temperature at which the zero-field cooled magnetization has its maximum decreases with an external magnetic field in line with the Gabay-Toulouse prediction. The temperature at which the AC magnetic susceptibility has its maximum does not depend on frequency which, in the light of the mean-field theory, testifies to very long magnetic interactions.

Key words:

 A. Intermetallics,

B. Solid State Reactions

C. Magnetization

D. Magnetic measurements

[*]Corresponding author: Stanislaw.Dubiel@fis.agh.edu.pl; fax: +48 126340010

## 1. Introduction

The tetragonal sigma phase ($\sigma$) (space group $D^{14}_{4h}$ - $P4_2/mnm$ ) belongs to a Frank-Kasper (FK) family phases [1]. It can be formed in alloys in which at least one of constituting element is a transition metal. Its unit cell hosts 30 atoms occupying 5 different lattice sites having high coordination numbers (12-16). These features and the fact that $\sigma$ can be formed in a certain range of composition, make it possible to tailor the physical properties of the alloys by changing constituting elements and/or their relative concentration. Structural complexity and chemical disorder make them furthermore an attractive yet challenging subject for investigations. The interest in $\sigma$ has been additionally stimulated by a deteriorating effect of $\sigma$, on useful properties of technologically important materials in which it has precipitated [2,3]. On the other hand, attempts have been undertaken to take advantage of its high hardness for materials strengthening purposes e. g. [4,5].

Concerning magnetic properties of $\sigma$ in binary alloys, until recently only $\sigma$ in Fe-Cr and Fe-V alloy systems was definitely evidenced to possess such properties which were termed as ferromagnetic ones [6-8]. Its magnetism was, however, lately shown to be more complex than initially anticipated viz. in both cases it has a re-entrant character [8]. Furthermore, nuclear magnetic resonance measurements performed on $\sigma$-Fe-V samples revealed that vanadium atoms present on all five sub lattices were magnetic [10]. Newly, the magnetism of $\sigma$ was found in Fe-Re [11] and in Fe-Mo [12,13] alloy systems with a spin-glass constituting the ground magnetic state in both cases.

In this Letter results obtained from DC and AC magnetic susceptibility measurements concerning magnetic properties of a $\sigma$-$Fe_{54}Cr_{46}$ sample are presented and discussed.

## 2. Experimental

### 2.1. Sample preparation and composition

A master alloy of nominal composition of $Fe_{54.2}Cr_{45.8}$ was prepared by melting the appropriate amounts of Fe (99.95% purity) and Cr (99.5% purity) in an arc furnace under a protective atmosphere of argon. This process of melting was repeated three times in order to ensure a better degree of chemical homogeneity. The ingot was next annealed in vacuum for 72 h at 1273 K and fast cooled outside the furnace. A plate of ~7x4x3 mm$^3$ cut out of the ingot, together with the rest of the ingot were transformed into σ by an isothermal annealing for 450h at 973K. **Powder** X-ray diffraction pattern recorded on the latter gave evidence that the alloy after the isothermal treatment was 100% σ. Energy dispersive spectroscopy (EDS) studies were measured by using the EDS-JOEL JSM-7700 scanning electron microscope. EDS chemical analysis shows a uniform distribution of 45.9±0.6 at% Cr concentration and also a small amount of oxygen in some points on the surface.

### 2.2. Magnetic susceptibility measurements

DC Magnetization (*M*) measurements at various applied magnetic fields (H) in the temperature interval 5 K < T < 300 K, as well as at different isothermal *M* versus *H* up to 50 kOe, have been performed using the commercial (Quantum Design) superconducting quantum interference device (SQUID) magnetometer with sample mounted in gel-caps. Prior to recording the zero-field-cooled (ZFC) curves, the SQUID magnetometer was always adjusted to be in a "true" H = 0 state. The temperature dependence of the field-cooled (FC) and the ZFC branches were taken via warming the samples. The real ($\chi'$) and imaginary ($\chi''$) AC magnetic susceptibilities (at *H*=0) were measured with a home-made pickup coil method at an AC field amplitude of $h_0$=0.05 Oe at frequencies of up to 1465 Hz. The AC measurements were performed by using the same SQUID magnetometer.

**3. Results and discussion**

**3.1. DC magnetic susceptibility**

Isothermal magnetization data recorded at various temperatures are displayed in Fig. 1. It is clear that a saturation has not been achieved which agrees with our previous measurements [7,8]. Extrapolation of the linear part of the data recorded at 5 K to $H$=0 yields an average magnetic moment $M_S$=14.8 emu/g which corresponds to $<\mu>_{Fe}$=0.26 $\mu_B$ per Fe atom. This value agrees with the literature data [7,8]. The extrapolated value obtained at 122 K (4.6 emu/g) is probably due to ~5% $Fe_3O_4$ (magnetite, $M_S$=96 emu/g) which is ferrimagnetically ordered up to 856 K.

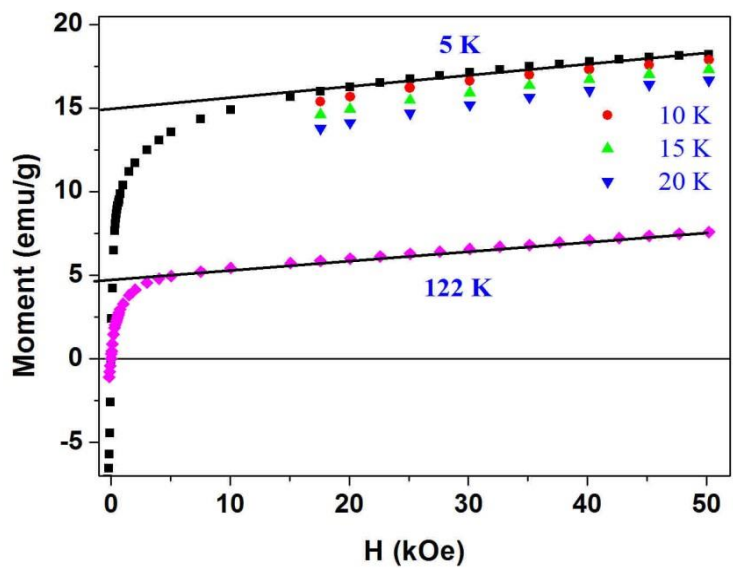

Fig. 1 Isothermal magnetization data recorded on the $\sigma$-$Fe_{54}Cr_{46}$ sample displayed versus external magnetic field, $H$. The solid line is the best fit to the linear parts of the data. Its intersection with the vertical axis yields the average magnetic moment per atom, ⟨µ⟩.

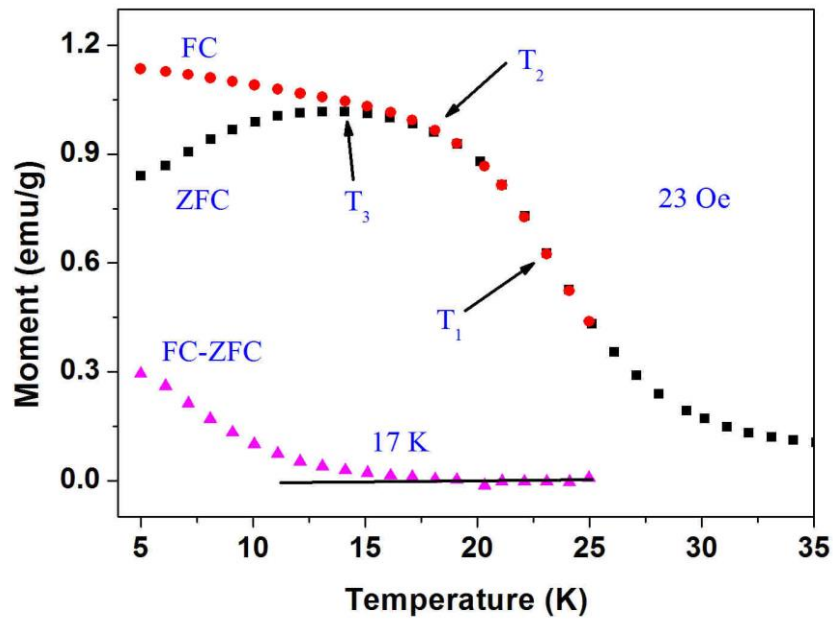

Fig. 2 Zero-field cooled (ZFC) and field-cooled (FC) magnetization data vs. temperature recorded at an external magnetic field of 23 Oe. The difference between the two sets of the data is displayed in the lower part of the plot showing the temperature of bifurcation, 17 K.

*ZFC and FC M(T)*-measurements were carried out in external magnetic fields of 23, 53 and 100 Oe. An example of the recorded data at ($H$=23 Oe) can be seen in Fig. 2. It clearly illustrates a behavior typical of a spin glass (SG) viz. a bifurcation effect. Three characteristic temperatures, $T_1$, $T_2$ and $T_3$, can be determined from the measurements. From the inflection point of the *FC*-curve (obtained by differentiation) one gets $T_1$ which can be regarded as the magnetic ordering temperature or the Curie point, $T_C$, the temperature at which the bifurcation occurs is indicated as $T_2$ and the maximum of the ZFC-curve determines $T_3$. The last two temperatures can be associated with the SG state. Namely, $T_2$ indicates a spin-freezing temperature while $T_3$ marks a transition from a weak into a strong irreversibility regime of SG. The obtained $T_k$-values are displayed as the H-T magnetic phase diagram in Fig. 3. It can be seen that whereas $T_1$ hardly depends on H both $T_2$ and $T_3$ decrease with H. Such a behavior agrees with theoretical models which predict the following relationship:

$$T_k(H) - T_k(0) = \Delta T_k \propto H^\phi \qquad (1)$$

Where $k=2,3$, $\phi = 2$ according to the Gabay and Toulouse (GT)-model [14] and $\phi = 2/3$ according to the de Almeida and Thouless (AT) model [15].

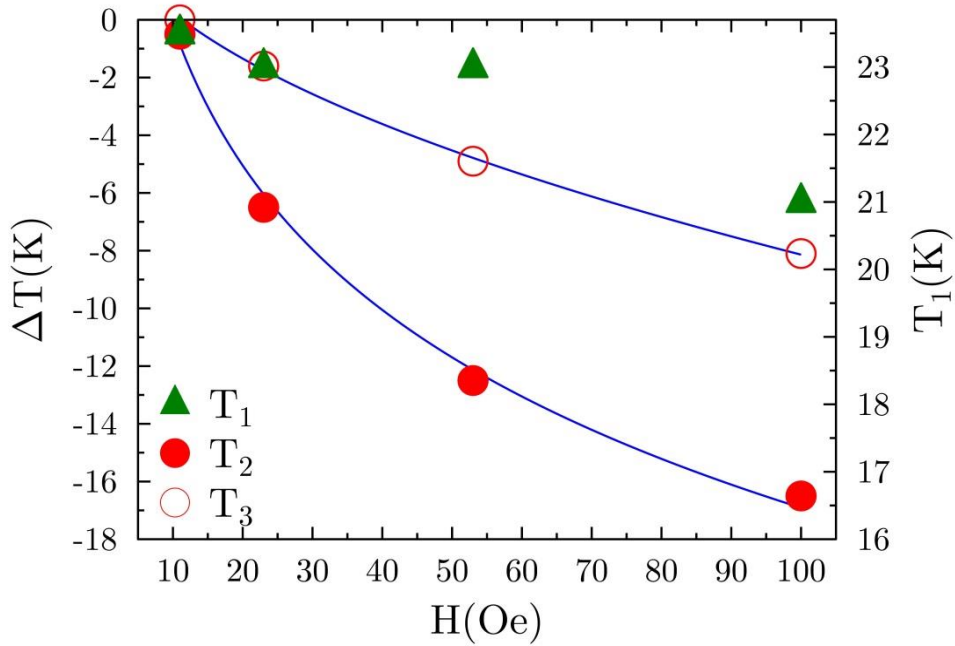

Fig. 3. Dependence of the temperatures $T_1$, $T_2$ and $T_3$ on external magnetic field, H. The lines represent the best-fits in terms of Eq. (2).

The analysis of the $T_k$-data ($k$=2,3) in terms of eq. (1) was successful only then when an $H$-independent term, $a$, was added i.e.

$$\Delta T_k = a + bH^\phi \qquad (2)$$

For $T_2$ the best-fit parameters were as follows: $a$= -140, $b$=155 and $\phi$ = 25 while for $T_3$ $a$ = -1.2, $b$=4 and $\phi$ = 2. The former result is unphysical, and it is likely due to the ill-defined bifurcation temperature while the latter agrees perfectly with the GT model.

The $M(T)$ data were also used to determine the effective paramagnetic moment, $\mu_{eff}$. Toward this end the magnetic susceptibility, $\chi = M/H$, was considered – see Fig. 4

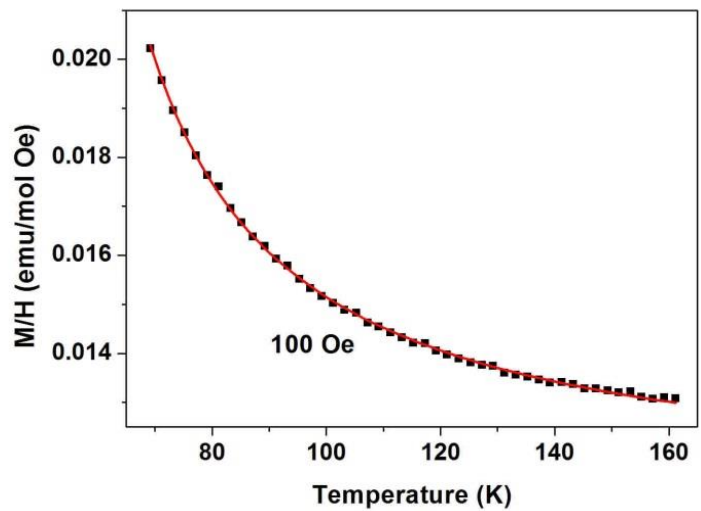

Fig. 4 The DC susceptibility, $\chi$=M/H, versus temperature as measured at $H$=100 Oe.

The susceptibility was fitted to the Curie–Weiss formula:

$$\chi = \chi_o + \frac{C}{T - \theta} \qquad (2)$$

Where C is the Curie constant and $\theta$ is the Curie-Weiss temperature.

The best-fit yields C=0.228(2) emu·K/mol·Oe which corresponds to $\mu_{eff}$=1.9 $\mu_B$/Fe atom, and a positive $\theta$ = 44.5 K, which indicates that in the paramagnetic state the effective interaction is ferromagnetic. The knowledge of $T_C$ ($T_1$) and $\theta$ enables determination of the degree of frustration, $FD = \theta/T_C \approx 2$, which testifies to a weak frustration.

### 3.2. AC magnetic susceptibility

As evidenced in Fig. 6, the real part of the susceptibility, $\chi'$, has a maximum that defines a spin-freezing temperature, $T_f$. Usually, $T_f$ depends on frequency, $f$, and the relative shift of $T_f$ per decade of frequency, RST, is used to make a distinction between different types of SGs: the smaller the RST-value the longer the range of magnetic interactions between magnetic moments. According to the mean field (MF) theory which assumes an infinite range of the interactions, RST=0 [16].

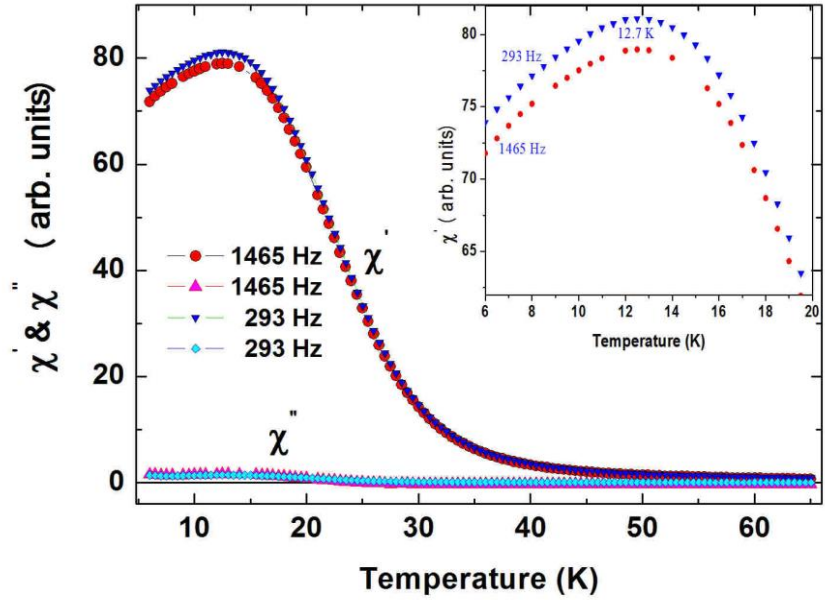

Fig. 5  Real, $\chi'$, and imaginary, $\chi''$, parts of the magnetic susceptibility recorded versus temperature for different frequencies shown. The inset shows the maxima for the two frequencies.

In our case $T_f$ =12.7 K and it does not depend on $f$. Consequently, $RST$=0 which, to our best knowledge, is the only known case so far where $T_f$ is frequency independent. In the light of the MF-theory it means that the range of the magnetic interactions responsible for the SG state is very long. The latter can be understood in terms of an itinerant character of magnetism in the studied sample which, in turn, follows from a very high value of the $\mu_{eff}/\langle\mu\rangle_{Fe}$ ratio viz. ~ 7. The latter, according to the Rhodes-Wohlfarth criterion, testifies to the band character of magnetism [17]. Interestingly, for the σ-phase Fe-Mo alloys that also show the itinerant magnetism, $RST$ was found to lie between 0.012 and 0.0135 [12]. Obviously, a higher degree of delocalization observed in the σ-phase Fe-Cr alloys than in the Fe-Mo ones is due to Cr which in a pure metallic phase is known to possess a very itinerant character of magnetism.

The inflection point determined from the $\chi'$-curve recorded for $f$=293 Hz is equal to 23.3 K, hence it perfectly agrees with the corresponding value found from the DC susceptibility measurements at 11 Oe.

The imaginary part of the AC susceptibility, $\chi''$, shows similar behavior, as shown in Fig. 7, i.e. its maximum occurs at 12.7 K and the inflection point (obtained by differentiation) is equal to 21K. In addition, it exhibits a minimum at 7.5 K whose origin is not clear.

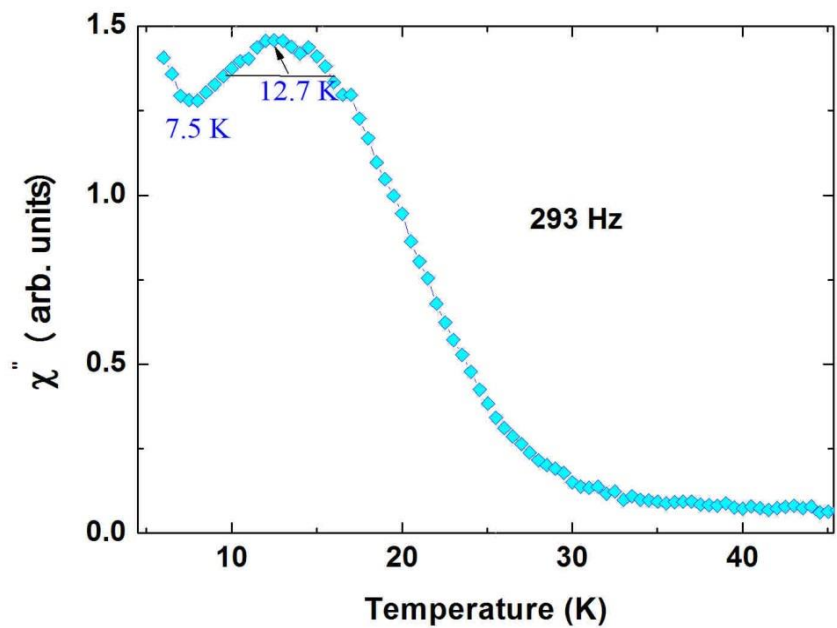

Fig. 7 Imaginary, $\chi''$, part of the magnetic susceptibility recorded versus temperature for $f$=293 Hz. The inset shows the derivative in the vicinity of the inflection point of 21K.

## 4. Conclusions

The following conclusions can be drawn from the results obtained in this study:

1. The investigated sample of $\sigma$-Fe$_{54}$Cr$_{46}$ is magnetic with the magnetic ordering temperature of 23.5 K

2. Its magnetism has a re-entrant character with a spin glass being the ground state.

3. The field dependence of the temperature at which the ZFC magnetization curve has its maximum is in line with the Gabay-Toulouse prediction.

4. The existence of the SG has been confirmed by the AC susceptibility measurements.

5. The spin-freezing temperature defined by the maximum in the real part of the AC susceptibility does not depend on frequency up to 1465 Hz which indicates a very long-range magnetic interactions.

## Acknowledgements

This study was supported by The Ministry of Science and Higher Education of The Polish Government.